\documentclass[showpacs,pra,superscriptaddress,notitlepage,twocolumn]{revtex4-2}
\usepackage{qcircuit}
\usepackage[dvips]{graphicx}
\usepackage{amsmath,amssymb,amsthm,mathrsfs,amsfonts,dsfont}
\usepackage{subfigure, epsfig}
\usepackage{braket}
\usepackage{bm}
\usepackage{fancyhdr}
\usepackage{multirow}
\usepackage{enumerate}
\usepackage{color}
\usepackage[dvipsnames]{xcolor}
\usepackage{lineno}

\usepackage{hyperref}
\hypersetup{colorlinks=true, linkcolor=blue, citecolor=blue, urlcolor=black }

\newcommand{\comments}[1]{}

\usepackage{comment}
\usepackage{graphicx}
\usepackage{amsmath}
\usepackage{amsfonts}
\usepackage{gensymb}
\usepackage{tabularx}
\usepackage{float}
\usepackage[linesnumbered, lined, ruled, boxed]{algorithm2e}

\SetKwRepeat{Do}{do}{while}
\newcommand{\rAA}{$\rm{\AA}$}

\newcommand{\bytedance}{ByteDance Inc, Zhonghang Plaza, No. 43, North 3rd Ring West Road, Haidian District, Beijing.}
\newcommand{\bytedanceus}{ByteDance Inc, Mountain View, CA 94041, US.}
\newcommand{\tsinghua}{Department of Chemistry, Tsinghua University,  Beijing 100084, China}

\newcommand{\peking}{Center on Frontiers of Computing Studies, Peking University, Beijing 100871, China}

\newcommand{\oxford}{Clarendon Laboratory, University of Oxford, Oxford OX1 3PU, UK}
\usepackage[utf8]{inputenc}

\usepackage{multirow}
\usepackage{booktabs}
\usepackage{comment}

\usepackage{soul}

\begin{document}

\title{\textit{Ab initio} simulation of complex solids on a quantum computer \\ with orbital-based multifragment density matrix embedding}

\author{Changsu Cao}
\affiliation{\bytedance}
\affiliation{\tsinghua}

\author{Jinzhao Sun}
\affiliation{\oxford}

\author{Xiao Yuan}
\affiliation{\peking}

\author{Han-Shi Hu}
\affiliation{\tsinghua}

\author{Hung Q. Pham}
\email{hung.pham@bytedance.com}
\affiliation{\bytedanceus}

\author{Dingshun Lv}
\email{lvdingshun@bytedance.com}
\affiliation{\bytedance}

\begin{abstract}
Quantum computing has shown great potential in various quantum chemical applications such as drug discovery, material design, and catalyst optimization. Although significant progress has been made in quantum simulation of simple molecules, \textit{ab initio} simulation of solid-state materials on quantum computers is still in its early stage, mostly owing to the fact that the system size quickly becomes prohibitively large when approaching the thermodynamic limit. In this work, we introduce an orbital-based multi-fragment approach on top of the periodic density matrix embedding theory, resulting in a significantly smaller problem size for the current near-term quantum computer. We demonstrate the accuracy and efficiency of our method compared with the conventional methodologies and experiments on solid-state systems with complex electronic structures. These include spin polarized states of a hydrogen chain~(1D-H), the equation of state of a boron nitride layer~(h-BN) as well as the magnetic ordering in nickel oxide~(NiO), a prototypical strongly correlated solid. Our results suggest that quantum embedding combined with a chemically intuitive fragmentation can greatly advance quantum simulation of realistic materials, thereby paving the way for solving important yet classically hard industrial problems on near-term quantum devices.
\end{abstract} 
\maketitle

\section{Introduction}
Quantum simulation for the ground state problem of chemical systems has been among the most promising applications of quantum computing~\cite{Cao2019, QuantChem_review1,peruzzo2014variational, kandala2017hardware, nam2020ground,arute2020hartree,Lee2022}. 
Although the practical quantum computing advantage of chemistry problems is still on the way, the development of efficient quantum algorithms for noisy intermediate-scale quantum~(NISQ) devices has been a rapidly growing field of quantum technology~\cite{preskill2018quantum}. Hybrid quantum-classical algorithms like variational quantum eigensolver~(VQE)~\cite{peruzzo2014variational} and its variants have been viewed as 
promising candidates that are compatible with NISQ devices for quantum simulation of chemical systems. Considerable progress centered around variational algorithms has been made in quantum simulations of molecular systems via both 
numerical~\cite{Theo_VQEUCCSD1,Theo_VQEUCCSD2,C2H4_28qubits,H24_VQE_hardware_efficient,tazhigulovSimulatingChallenging2022} and experimental explorations~\cite{peruzzo2014variational,boyn2021quantum, smart2022resolving,kandala2017hardware,nam2020ground,arute2020hartree,huggins2022unbiasing}. Yet the experimental capability is no more than 20 qubits.

Despite the progress in molecular systems, algorithmic development in quantum simulation of materials is still in the early stage. 
Such studies could greatly enhance the capability of a quantum computer in solving both fundamental and industrial challenges, such as the understanding of unconventional superconductivity~\cite{HighT_superconductivity1} and the optimization of heterogeneous catalysis~\cite{deutschmann2013modeling,Bruix2019}. Nonetheless, quantum simulation of periodic systems remains challenging in the foreseeable future. The primary challenge to \textit{ab initio} simulation of periodic systems on a quantum computer has its root in the necessity of approaching the thermodynamic limit~(TDL) for quantitative predictions. 
This results in an extra dimension of computational variables compared to their molecular counterparts, thus demanding an impractically large number of qubits. So far, most of the current studies have been limited to toy models of simple solids like the hydrogen chain with a minimal basis set~\cite{yamamoto2021quantum,fan2021equation,liu2020simulating,manrique2020momentum, mizuta2021deep,yoshiokaVariationalQuantum2022}.

To tackle this problem and enable NISQ devices for quantum simulation of solid-state materials, various multiscale hybrid quantum-classical methods have been proposed~\cite{QuantChem_review1}. Quantum embedding theory has been the most widely used framework for such a multiscale simulation of chemical systems. In principle, a quantum embedding theory divides a large system into smaller fragments, and enables an accurate treatment using a high-level computational method. By integrating with a quantum solver like quantum phase estimation or VQE, quantum embedding allows the representation of the fragment on a NISQ device and the quantum subsystem is usually coupled to a classical environment. Fruitful progress has been made on quantum simulation of lattice models and small molecules within the framework of quantum embedding, for instance, dynamical mean-field theory~(DMFT)~\cite{bauerHybridQuantumClassicalApproach2016, runggerDynamicalMeanField2020, chenVariationalQuantumEigensolver2021, keenQuantumclassicalSimulationTwosite2020, jaderbergMinimumHardwareRequirements2020}, density matrix embedding theory~(DMET)~\cite{rubin2016hybrid_dmetqc1, kawashimaOptimizingElectronicStructure2021, tillyReducedDensityMatrix2021,DMET_QC_C18, greene-dinizModellingCarbonCapture2022}, and projection-based embedding~\cite{ralliScalableApproachQuantum2022}. 
Thus far, the pioneering attempts of quantum simulations on periodic solids have been focused on the active space \textit{ansatz} where the effective Hamiltonian is constructed using an active subspace of orbitals embedded in a Hartree-Fock~(HF) or density functional theory~(DFT) mean-field~\cite{maQuantumSimulationsMaterials2020,huangSimulatingElectronic2022,vorwerkQuantumEmbedding2022}. Although such an active space approach is a natural choice for systems where an active site can be straightforwardly selected, for example, spin defect in solids~\cite{maFirstprinciplesStudies2020,maQuantumEmbedding2021,shengGreenFunction2022}, a general framework for \textit{ab initio} quantum simulation of solid-state materials is still missing. We are particularly interested in DMET as a promising platform to derive an efficient hybrid quantum-classical algorithm for two reasons. DMET intrinsically allows high-level treatments for multiple fragments or correlated sites using a high-level theory unlike the active space or projection-based embedding which have been mostly developed for one single fragment or one correlated site (for instance, solid-state defect~\cite{mitra2021excited}). Additionally, DMET is a computationally cheaper alternative to the successful DMFT technique~\cite{DMET_Knizia1} in order to accelerate the convergence of the ground-state energy to TDL.

While recent developments in \textit{ab initio} DMET~\cite{DMET_Knizia1, DMET_Knizia2, DMET_Wouters, pDMET_Hung, pDMET_Cui} and its multifragment extension have been applied for periodic solids~\cite{cui2021systematic}, the fragment size can be hundreds of qubits or more, which are hard to solve classically or using NISQ hardware.
Here, by partitioning the unit cells into multiple subsets of strongly and weakly correlated orbitals, we introduce a hybrid quantum-classical algorithm that could overcome the limitation.
The embedding Hamiltonian is now solved in a hybrid manner, where the strongly correlated subsets are solved by a quantum solver and the weakly correlated subsets are treated classically. This orbital-based multifragment approach is motivated by the observation that strong correlation in materials can be attributed to a few correlated orbitals, for example, 3$d$ orbitals in transition metal oxides or cuprates~\cite{Kotliar2004}. The strongly correlated subsets are often small and well suited given the current limitation of quantum resources. We investigate the performance of our method using prototypical systems from weakly to strongly correlated electronic structure, including the spin polarization of one-dimensional hydrogen chain~(1D-H), equation of state~(EOS) of two-dimensional hexagonal boron nitride~(h-BN), as well as the magnetic ordering of three-dimensional nickel oxide~(NiO). We show that the quantum resource required for the largest simulation on NiO can be reduced to 20 qubits from requirement of 9984 qubits using a non-embedding algorithm. Overall, our results suggest that it is possible to perform quantitative prediction on realistic solids using quantum computers. This work 
paves the way for further researches in \textit{ab initio} simulation of large and complex periodic systems on near-term quantum devices. 

\section{Framework}
In this section, we introduce the key steps in the \textit{ab initio} periodic DMET algorithm used in this work. We discuss how this algorithm can be integrated with an orbital-based partition of the unit cell to enable the simultaneous use of both quantum and classical solvers, resulting in a hybrid quantum-classical algorithm for realistic solids. 
For periodic systems, each atomic Bloch orbital, or a \textbf{k}-adapted atomic orbitals (AOs), is simply the Fourier transform of a contracted Gaussian-type orbital (cGTO):
\begin{equation}
\phi^{\textbf{k}}_{\mu}(\textbf{r})=\sum_{\textbf{R}}e^{\rm{i}\textbf{kR}}\phi_{\mu}(\textbf{r}-\textbf{R}),
\label{kBasis}
\end{equation}
where \textbf{R} is a real space lattice vector and \textbf{k} is a crystal momentum. The crystalline orbitals are linear combinations of atomic Bloch orbitals, $\psi^{\textbf{k}}_p(\textbf{r}) = \sum_{\mu} C^{\textbf{k}}_{\mu p} \phi^{\textbf{k}}_{\mu}(\textbf{r})$. The second-quantized Hamiltonian in the crystalline orbital is expressed as
\begin{equation}
\begin{split}
\hat{H} = &\sum_{p,q}\sum_{\textbf{k}_p \textbf{k}_q} {h^{\textbf{k}_p \textbf{k}_q}_{pq}\hat{a}^{\textbf{k}_p\dagger}_p \hat{a}^{\textbf{k}_q}_q } \\ & + \frac{1}{2} \sum_{p,q,r,s} \sum_{\textbf{k}_p \textbf{k}_q \textbf{k}_r \textbf{k}_s} {g^{\textbf{k}_p \textbf{k}_q \textbf{k}_r \textbf{k}_s}_{pqrs} \hat{a}^{\textbf{k}_p\dagger}_p \hat{a}^{\textbf{k}_r\dagger}_r \hat{a}^{\textbf{k}_s}_s \hat{a}^{\textbf{k}_q}_q  }.
\label{kHam1}
\end{split}
\end{equation}
where $\hat{a}^{\textbf{k}_p\dagger}_p$~($\hat{a}^{\textbf{k}_p}_p$) is the fermionic creation~(annihilation) operator to $p$-th orbital in $\textbf{k}_p$ sampled in momentum space~(\textbf{k}-space), with $p$, $q$, $r$, $s$ denoting the indices of crystalline orbitals.
\textit{Ab initio} simulation on solids usually requires the computation on a supercell with a very large size. Thus, it is a common practice to employ the transnational symmetry by solving the Hamiltonian in the momentum space~(\textbf{k}-space). In Eq.~\eqref{kHam1}, we note that $\textbf{k}_p - \textbf{k}_q = 0 $ and $\textbf{k}_p + \textbf{k}_r - \textbf{k}_s - \textbf{k}_q = 0$ owing to the translational symmetry (this is preferred as the conservation of crystal momentum).
In the following, the localized orbitals are denoted by $i$, $j$, $k$ and $l$ while the orbitals in embedding space are denoted by $\tilde{i}$, $\tilde{j}$, $\tilde{k}$ and $\tilde{l}$.

\begin{figure*}[ht!]
\centering
  \includegraphics[width=.85\linewidth]{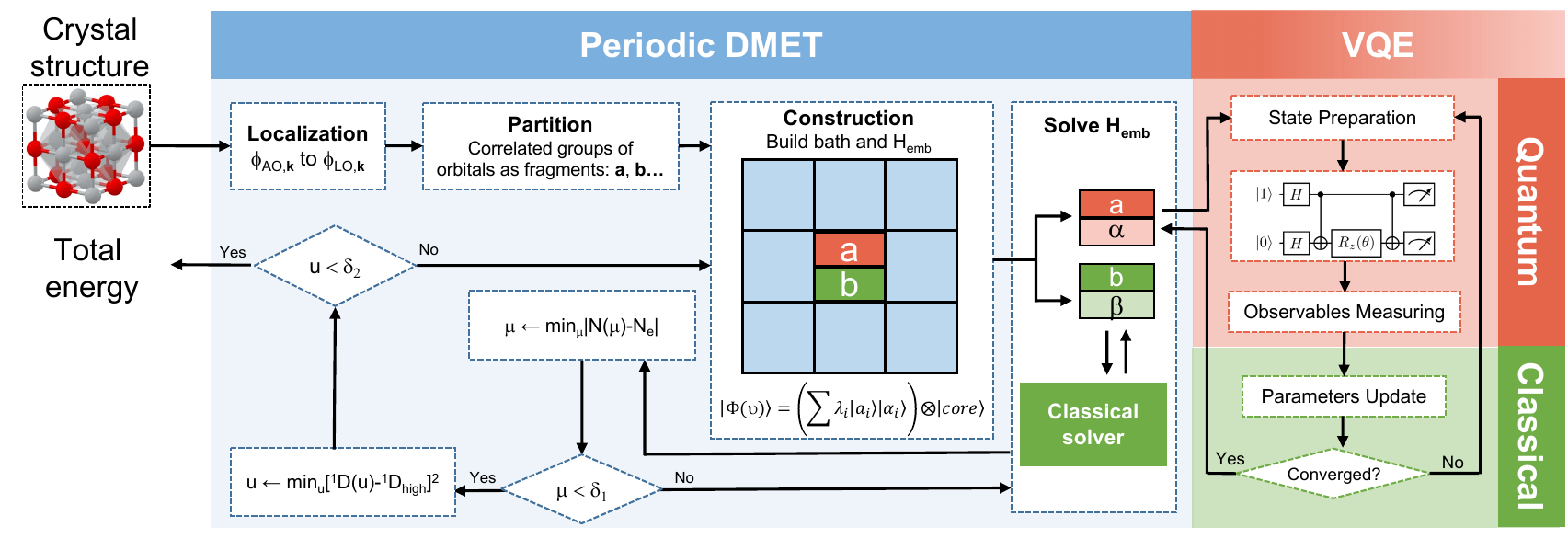}
  \caption{Schematic diagram of the hybrid quantum-classical periodic DMET algorithm for the \textit{ab initio} simulation of solids. In this cartoon, we assume the unit cell is divided to two fragments \textbf{a} and \textbf{b} without loss of generality. Due to the translational symmetry in the periodic system, we only perform the partition covering one unit cell and solve them in high-level solver without repeating the calculations on all the other unit cells.
  }
  \label{fig:VQE-DMET}
\end{figure*}

The quantum-classical hybrid flowchart is illustrated in Fig.~\ref{fig:VQE-DMET}. 
The algorithm begins with the procedure of 
projecting the non-orthogonal atomic Bloch orbitals to the orthogonal localized ones, $i.e.$ periodic intrinsic atomic orbitals~(IAO)~\cite{IAO} and projected atomic orbitals~\cite{PAO}, and partitioning the unit cell into several correlated fragments.
In previous DMET algorithms for periodic systems, the whole unit cell is usually selected as one unique fragment embedded in an environment composed of unit cells~\cite{pDMET_Hung, pDMET_Cui}. Due to the translational symmetry, only one embedding problem needs to be solved. It has also been extended to multi-fragments in multi-layer compounds or inhomogeneous systems, where a sub-structure containing several atoms is chosen as the fragment~\cite{cui2021systematic}. In this work, we introduce an orbital-based fragmentation of which each fragment may contain several correlated orbitals from different atoms. Within this scheme, it is possible to perform the partition with arbitrary number of fragments. For the sake of simplicity, we assume the unit cell is divided into two fragments, \textbf{a} and \textbf{b}, to be solved with quantum and classical solvers, respectively.

The bath of each fragment is constructed solely by treating the rest of the supercell as the environment in \textbf{R}-space. 
For fragment \textbf{a}, it is built by performing singular value decomposition on the off-diagonal block of density matrix between the fragment and the environment, which is expressed as, 

\begin{equation}
D_{ij}= \sum_{ \tilde{k}}B_{i \tilde{k}} \lambda_{\tilde{k}} V_{\tilde{k} j}^{\dagger} \text{ with } i \subset \textbf{a}, j \not\subset \textbf{a}
\label{Bath_SVD}
\end{equation}
where $B$ is a rectangular matrix used to transform the local orbitals to the bath orbitals. The embedding orbitals are composed of the fragment and the bath orbitals. The number of the bath orbitals is no larger than that of the fragment. Regardless of the size of the supercell, the size of each embedding problem is at most $2N$ with $N$ is the number of local orbitals within the fragment. Then we obtain the embedding Hamiltonian similarly to molecular systems through a projector operator~(which is described in detail in the literature~\cite{DMET_Wouters,pDMET_Cui,DMET_QC_C18}) as follows
\begin{equation}
      \hat{H}_{\text{emb}}=
      \sum_{\tilde{i},\tilde{j}} {h}_{\tilde{i} \tilde{j}} \hat{a}_{\tilde{i}}^{\dagger} \hat{a}_{\tilde{j}}
      - \mu \sum_{\tilde{i} \subset \textbf{a}} \hat{a}_{\tilde{i}}^{\dagger} \hat{a}_{\tilde{i}}
      +\frac{1}{2}\sum\limits_{\tilde{i},\tilde{j},\tilde{k},\tilde{l}} g_{\tilde{i} \tilde{j} \tilde{k} \tilde{l}}\hat{a}_{\tilde{i}}^{\dagger}\hat{a}_{\tilde{k}}^{\dagger}\hat{a}_{\tilde{l}} \hat{a}_{\tilde{j}},
\label{H_emb}
\end{equation}
where $\hat{a}_{\tilde{i}}^{\dagger}$ and $\hat{a}_{\tilde{j}}$ are operators that create and annihilate electrons in the embedding orbital $\tilde{i}$ and $\tilde{j}$, respectively. $h_{\tilde{i}\tilde{j}}$ and $g_{\tilde{i} \tilde{j} \tilde{k} \tilde{l}}$ are the one- and two-body component of the embedding Hamiltonian which are evaluated classically; and $\mu$ is the global chemical potential, which is a parameter to be optimized in DMET procedures to ensure the conservation of electron numbers.

Once constructed, a correlated wave function method is typically employed to solve the embedding Hamiltonian. The key difference is we only utilize the quantum resources for the embedding problem corresponding to the strongly correlated group while a classical solver, for instance, coupled cluster singles and doubles~(CCSD) is used for the weakly correlated group. The size of the strongly correlated group orbitals can be managed to accommodate the current as well as any future NISQ devices. In our implementation, the embedding Hamiltonian $H_{\text{emb}}$ for fragment \textbf{a} is passed to the quantum device and solved by the VQE with a spin-unrestricted unitary CCSD~(UCCSD) \textit{ansatz}~\cite{BARTLETT1989133,doi:10.1063/1.5011033,Taube2006}. {It is important to note that UCCSD has superior performance compared to traditional CCSD. This is due to its increased expressibility, as it includes both excitation and de-excitation operators. Additionally, UCCSD is more robust and avoids the non-variational catastrophe that often occurs in bond-breaking regions, unlike CCSD which tends to fail in these scenarios\mbox{\cite{harshaDifferenceVariational2018a,leeGeneralizedUnitary2019}}.}
The central idea of using VQE as the DMET solver is to construct and measure a parameterized embedding quantum state $\Psi(\vec{\theta})_{\text{emb}}$ on a quantum computer (see Fig.~\ref{fig:VQE-DMET}). For a UCCSD \textit{ansatz}, the parameter $\Vec{\theta}$ represents the amplitude for an excitation.
After obtaining the expectation value $ {\langle\Psi(\vec{\theta})_{\text{emb}}|\hat{H}_{\text{emb}}|\Psi(\vec{\theta})_{\text{emb}}\rangle}$ of embedding Hamiltonian, the amplitudes are updated on a classical computer variationally and iteratively until reach the terminal condition.
The quantum state constructed from UCCSD is expressed as $|\Psi(\vec{\theta})_{\text{emb}}\rangle = \exp{(\hat{T}(\vec{\theta}) - \hat{T}^{\dagger}(\vec{\theta}))}|\Psi_{0}\rangle$, where $|\Psi_{0}\rangle$ is the Hartree-Fock ground state of the embedding in the periodic system and $\hat{T}$ represents the summation of single excitations and double-excitations and defined as,

\begin{equation}
\begin{split}
\hat{T}(\boldsymbol{\theta}) &= \sum\limits_{\tilde{m},\tilde{a}, \sigma }\hat{t}_{\tilde{m} \sigma}^{\tilde{a} \sigma} + \sum\limits_{\tilde{m},\tilde{n},\tilde{a},\tilde{b},\sigma,
\tau}\hat{t}_{\tilde{m} \sigma, \tilde{n} \tau}^{\tilde{a} \sigma, \tilde{b} \tau} \\
&= \sum\limits_{\tilde{m},\tilde{a},\sigma}\theta_{\tilde{m} \sigma}^{\tilde{a} \sigma}\hat{a}_{\tilde{a} \sigma}^{\dagger}\hat{a}_{\tilde{m} \sigma}
 + \sum\limits_{\tilde{m},\tilde{n},\tilde{a},\tilde{b},\sigma, \tau} \theta_{\tilde{m}\sigma, \tilde{n} \tau}^{\tilde{a} \sigma, \tilde{b} \tau}\hat{a}_{\tilde{a} \sigma}^{\dagger}\hat{a}_{\tilde{b} \tau}^{\dagger}\hat{a}_{\tilde{n} \tau} \hat{a}_{\tilde{m} \sigma},
\label{cc_opt}
\end{split}
\end{equation}
where the occupied orbitals are denoted by $\tilde{m}$ and $\tilde{n}$ while the virtual orbitals are denoted by $\tilde{a}$ and $\tilde{b}$; $\sigma$ and $\tau$ are the corresponding spin indices. For an unrestricted UCCSD, occupation and excitation amplitude of the spin-up and spin-down orbitals are allowed to be different. The converged quantum circuit with variationally optimal parameters will be used to measure the one- and two-particle reduced density matrix, \textit{i.e.} $^{1}D^{\text{\text{QC}}}_{\text{emb}}$ and $^{2}D^{\text{\text{QC}}}_{\text{emb}}$, respectively; Similar to the classical algorithm these quantities are required to calculate the energy contribution from each fragment as well as to optimize a global chemical potential. 

Once we solve the embedding Hamiltonian, a correlation potential $u$ is employed to minimizes the difference of one-particle density matrix between the correlated density matrix $^{1}D^{\text{\text{QC}}}_{\text{emb}}$ and the mean-field~(mf) density matrix $^{1}D^{\rm{mf}}_{\text{emb}}$. 
This is often defined as a self-consistent DMET, denoted in this work as sc-DMET, to distinguish itself with a one-shot DMET~(o-DMET) where the mean-field bath is not optimized. In our algorithm, we locally search for the sub-block of the correlation potential or $u_{\textbf{a}}$ only for the strongly correlated fragment \textbf{a} while leave the weakly correlated fragment \textbf{b} aside.

\begin{equation}
\hat{F}'_{i j}=
\left\{
\begin{aligned}
& \hat{F}_{i j} + u_{i j} \hat{a}^{\dagger}_{i}\hat{a}_{j} & \text{ with } i \subset \textbf{a} \text{ and } j \subset \textbf{a}  \\
& \hat{F}_{i j}  & \text{ with } i \not\subset \textbf{a} \text{ or } j \not\subset \textbf{a} \text{      }
\end{aligned}
\right.
\label{fock_opt}
\end{equation}
The cost function thus can be written as
\begin{equation}
L(u_{\textbf{a}}) = \sum_{\tilde{p},\tilde{q} \subset \textbf{a}} {^{1}D_{\tilde{p} \tilde{q}}^{\text{classical}}(\hat{F}_{\tilde{p} \tilde{q}}(u)) - ^{1}D_{\tilde{p} \tilde{q}}^{\text{\text{QC}}}},
\label{cost_func}
\end{equation}
where $^{1}D_{\tilde{p} \tilde{q}}^{\text{classical}}$ is the mean-field one-particle reduced density matrix corresponding to the total system. 
Note if there is more than one strongly correlated fragments, the global correlation potential $u$ will be reconstructed by simply concatenating all submatrices $u_{\textbf{a}}$ into the corresponding submatrix. From our numerical experiments below, we have observed that this strategy usually leads to a fast convergence of the correlation potential with a satisfactory accuracy. Note that our proposed algorithm can be directly applied to multiple fragments, thus enabling a flexible framework to combine different quantum and classical solvers for the electronic structure of one crystal.

Once the self-consistent iteration terminates, the physical property of interest, such as total energy of the unit cell can be obtained through a sum over the contributions from different fragments, \textit{i.e.} $E_{\text{cell}} = E_a + E_b$, in a democratic manner that has been discussed several times in the literature~\cite{Wouters16}.

{Lastly, we discuss how our method takes into account the interaction between fragments through a self-consistent procedure. In particular, the correlation potential is found by optimizing a cost function that depends on the many-body reduced density matrix $^{1}D_{\tilde{p} \tilde{q}}^{\text{\text{QC}}}$. Additionally, the effective Hartree-Fock potential is reconstructed at each iteration by reusing the previous global many-body reduced density matrix which is discussed in previous work{~\cite{wuProjectedDensity2019, cui2021systematic}}, and is included in the one-body component of the embedding Hamiltonian. We assert that the many-body effect between the fragments is implicitly captured during the DMET's self-consistent procedure, resulting in partial inclusion of the many-body screening effect in our calculations. However, it is worth noting that the screening effect can be fully captured by systematically including more orbitals in the correlated fragments. Our orbital-partitioned extension of DMET possesses a high degree of generality and adaptability, allowing for such systematic convergence to the fully-screening effect without overwhelming the quantum resources. As such, it is promising in the NISQ era where the number of available qubits is gradually increasing over time.}

\section{Results and discussion}
\subsubsection{Spin polarized states of the hydrogen chain.}
\begin{figure}[ht]
\centering
  \includegraphics[width=\linewidth]{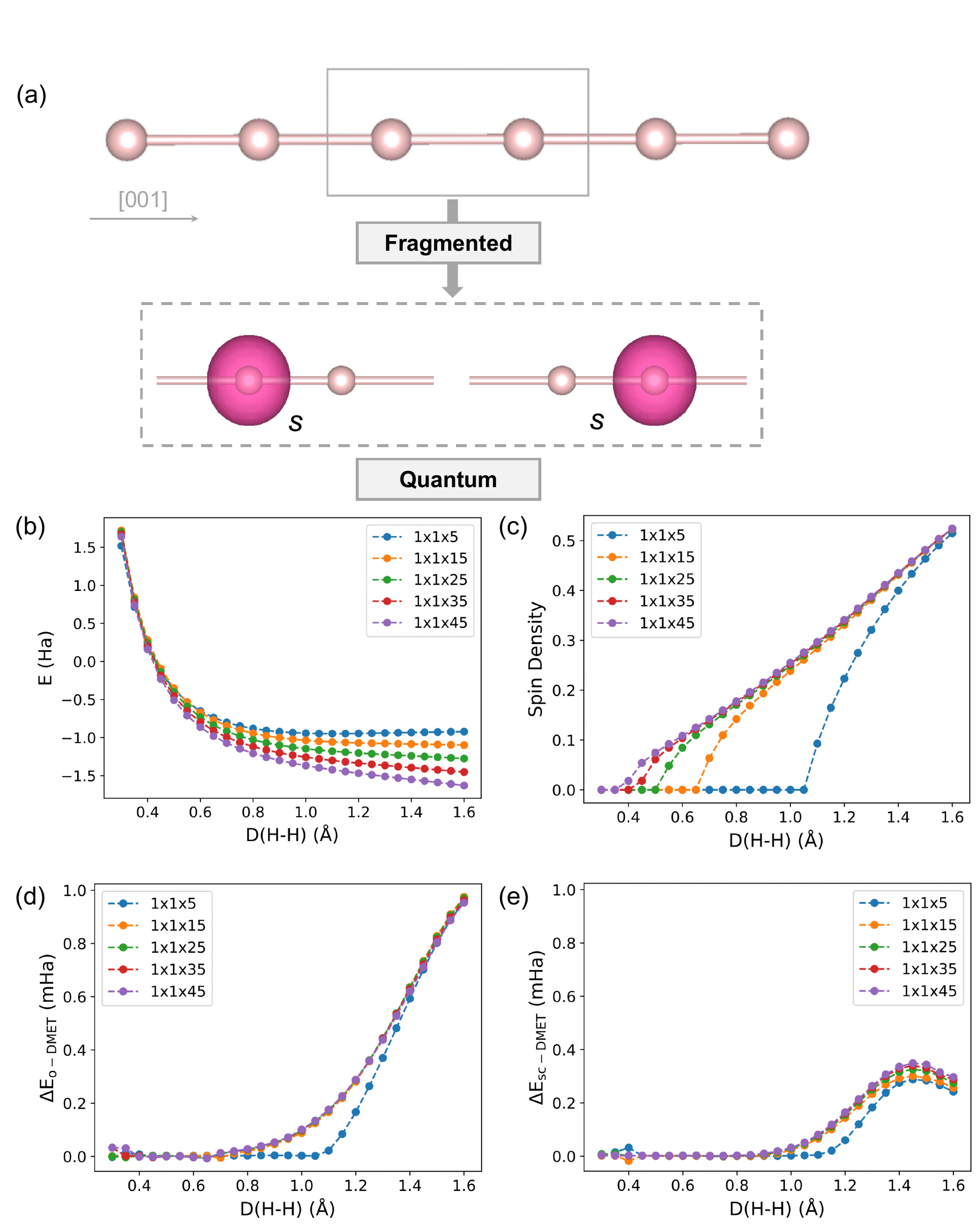}
  \caption{(a) The unit cell of 1D-H containing two hydrogen atoms is treated as the fragment in DMET calculations. (b) {The potential energy curve} of 1D-H calculated using different \textbf{k}-meshes by sc-DMET using QC solver. (c) The spin density, $i.e.$ the difference between the spin-up and spin-down electron density, using the Mulliken population~\cite{mullikenElectronicPopulation1955} for the hydrogen atom which shows the phase transition from the AFM state to the non-spin polarized state. (d) o-DMET and (e) sc-DMET energy differences between QC solvers and classical FCI solvers. The same scale for vertical axis is used.}
  \label{fig:HChain}
\end{figure}

We first perform numerical simulations on the one-dimensional equidistant hydrogen chain~(1D-H). As a simple but useful model, 1D-H is regarded as a good starting point to understand more complex systems~\cite{sinitskiyStrongCorrelation2010,stellaStrongElectronic2011,SolutionManyElectron2017}. In this part, we use a spin-unrestricted UCCSD \textit{ansatz} in VQE as quantum computing~(QC) solver integrated with DMET scheme to study the spin polarizability of 1D-H. 
{Both antiferromagnetic and non-spin polarized initial guesses were used to determine the non-magnetic ground-state with the lowest energy. The potential energy curve of the ground-state is depicted in the provided in Fig.{~\ref{fig:HChain}}(b).}
In our calculations, each unit cell containing two hydrogen atoms is treated as the fragment~(defined as a unit cell embedding throughout this work).

{In Fig.~{\ref{fig:HChain}}(b) and (c) show the total energy and spin-density of 1D-H calculated by QC solver~(QC-DMET). As the distance between two hydrogen atoms increases, the energy curves become flat and the spin density on the hydrogen atom increases}. This indicates that the system acts more like isolated atoms with a weak coupling between the hydrogen atoms. As the system is compressed, the spin polarization vanishes. {The transition from antiferromagnetic to non-spin polarized occurs approximately at 0.45 \AA~for the 1$\times$1$\times$45 \textbf{k}-mesh.  We have also compared our finding to those of previous studies and found them to be consistent, as detailed in section II of electronic supplementary information~(ESI).}

{We assess the accuracy of our method by comparing the DMET energies obtained with QC solvers to those from fully classical calculations. As shown in Fig.~{\ref{fig:HChain}}(d), the difference between the total energies calculated by o-DMET using QC solver and FCI solver is negligible for the compressed region where D(H-H) $\leq$ 1.0 \AA. The disparity between classical and quantum  solver increases as the hydrogen chain is stretched. It is worth noting that the sc-DMET method performs better than the o-DMET method in all scenarios as shown in Fig.~{\ref{fig:HChain}}(e). In particular, the results of sc-DMET exhibit a smaller classical-quantum deviation compared to those of o-DMET. Further analysis of the energy difference between the solvers can be found in section II in ESI.}

{Lastly, we also conducted simulations on the closed-shell 1D-H to investigate the effect of noise on our DMET algorithm. Our results indicate that the o-DMET energy converges with increasing number of shots and is resistant to depolarizing noise. Quantum error mitigation strategies were found to improve performance at moderate levels of noise. Additional information can be found in section V of ESI.}

\subsubsection{Equation of state of 2D h-BN.~~}
\begin{figure}[htp] 
\centering
  \includegraphics[width=\linewidth]{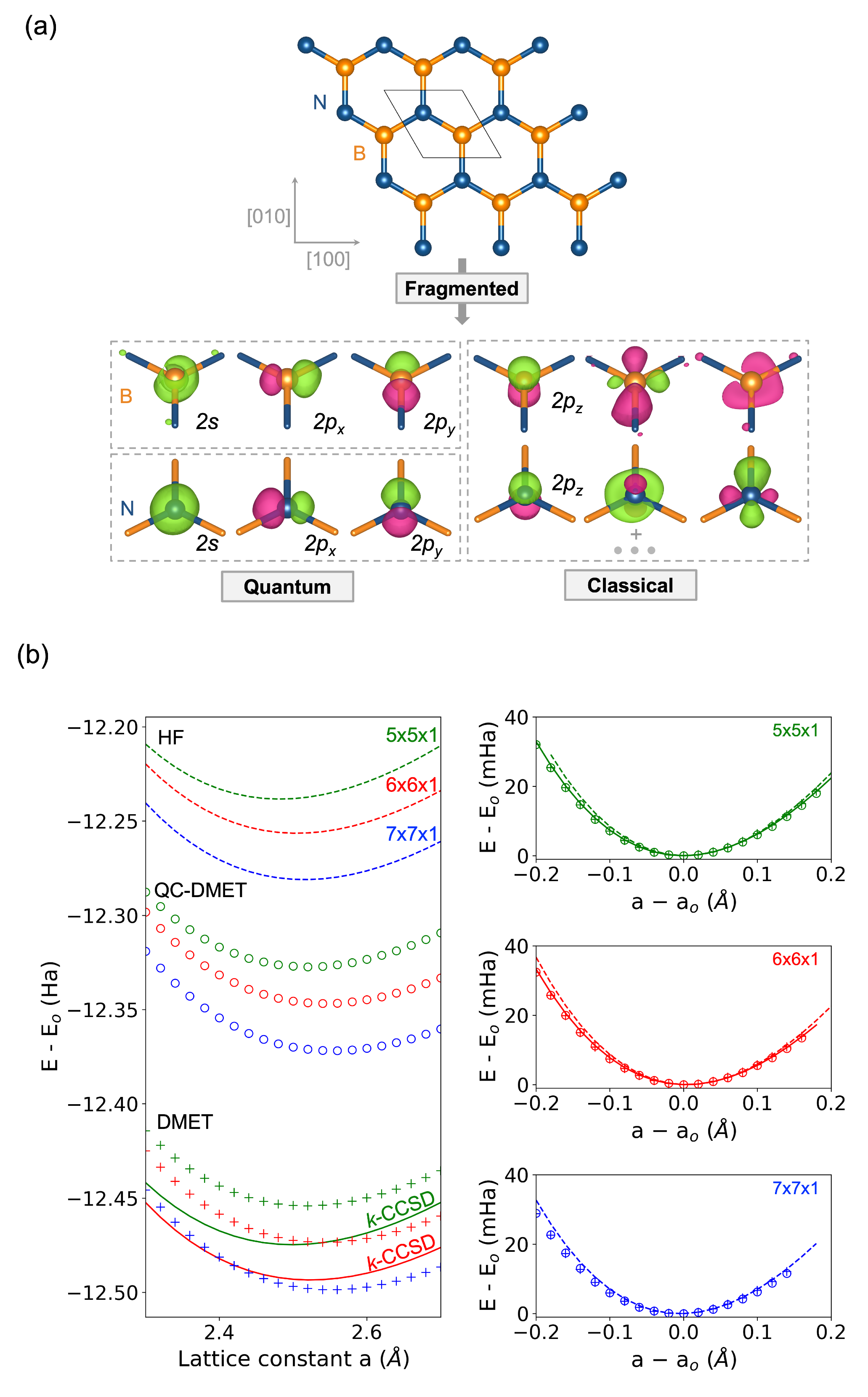}
   \caption{(a) The unit cell of BN is fragmented into three fragments and their corresponding IAOs. The first fragment is the $2s$, $2p_x$, and $2p_y$ orbitals of B while the second fragment is the same orbitals of N. The third fragment is the $p_z$ orbitals from B or N and all other orbitals; only several of them are shown here. (b) {EOS of BN by Hartree-Fock (dashed curves), classical unit cell DMET (plus signs), QC-DMET (unfilled circles), and {\bf k}-CCSD (solid curves) at different meshes of {\bf k}-points}: the absolute EOS (left) and the shifted EOS (right). The shifted EOS for each method was obtained by shifting both the minimal energy and the lattice constant corresponding to this minimum to zero. The 5$\times$5$\times$1, 6$\times$6$\times$1, and 7$\times$7$\times$1 {\bf k}-mesh are denoted by green, red, blue, respectively.}
  \label{fig:BN} 
\end{figure}    
 We investigate the performance of DMET with the orbital-based partition in calculating the EOS of a prototypical 2D system, the hexagonal boron nitride~(h-BN)~\cite{BN2011} with weak electron correlation. We note that this system has been studied before with a unit cell or multiple unit cells as the fragments~\cite{pDMET_Cui}. Inspired by the $sp^2$ hybridization of B and N~\cite{Bu2020}, we divide the unit cell of h-BN into three fragments (see Fig.~\ref{fig:BN} (a)). The first fragment consists of the $2s$, $2p_x$, and $2p_y$ orbitals of B while the second fragment consists of similar orbitals of N. The third fragment consists of all remaining orbitals. Since each of the first two fragments has only six spin orbitals, this allows us to utilize a QC solver while employing classical solvers for the third fragment. DMET using one unit cell as well as \textbf{k}-CCSD calculations are also performed for the reference.

Fig.~{\ref{fig:BN}} (b) presents the EOS of h-BN using different meshes for \textbf{k}-mesh. It can be observed that the orbital-partitioned DMET scheme in conjunction with the QC solver underestimates the absolute correlation energy when compared to \textbf{k}-CCSD. However, the computed EOS by all three methods exhibit a similar curvature, indicating that the orbital-partitioned scheme accurately captures the curvature of the EOS. This trend is consistent across different \textbf{k}-meshes. It is worth noting that the equilibrium lattice constant is a more relevant observable to benchmark a method as opposed to the curvature of the EOS. However, determining the lattice constant from an embedding technique necessitates a systematic convergence of the EOS with regards to the size of the fragment~{\cite{pDMET_Cui}}. 
Nonetheless, these calculations demand a much larger number of qubits which is beyond our limitation in computational resources. Therefore, we limit our scope by using the curvature of the EOS to benchmark our quantum calculations against k-CCSD. Our results showed that the agreement between QC-DMET, classical DMET, and k-CCSD is excellent, whereas Hartree-Fock (HF) has a small discrepancy of 4 mHa at $-$0.2\rAA.

We anticipate that the orbital-based embedding approach combined with the chemical intuition can be helpful in predicting reaction barrier or phase transition in solid-state materials. Importantly, the size of the embedding Hamiltonian within this approach is relatively small yet can be systematically expanded by including basis functions with high angular momentum numbers. This allows a simple way to apply the current noisy quantum solvers to realistic chemical systems at medium- or large-size basis sets.

\subsubsection{Magnetic ordering in NiO.~~}
\begin{figure}[t]
\centering
  \includegraphics[width=\linewidth]{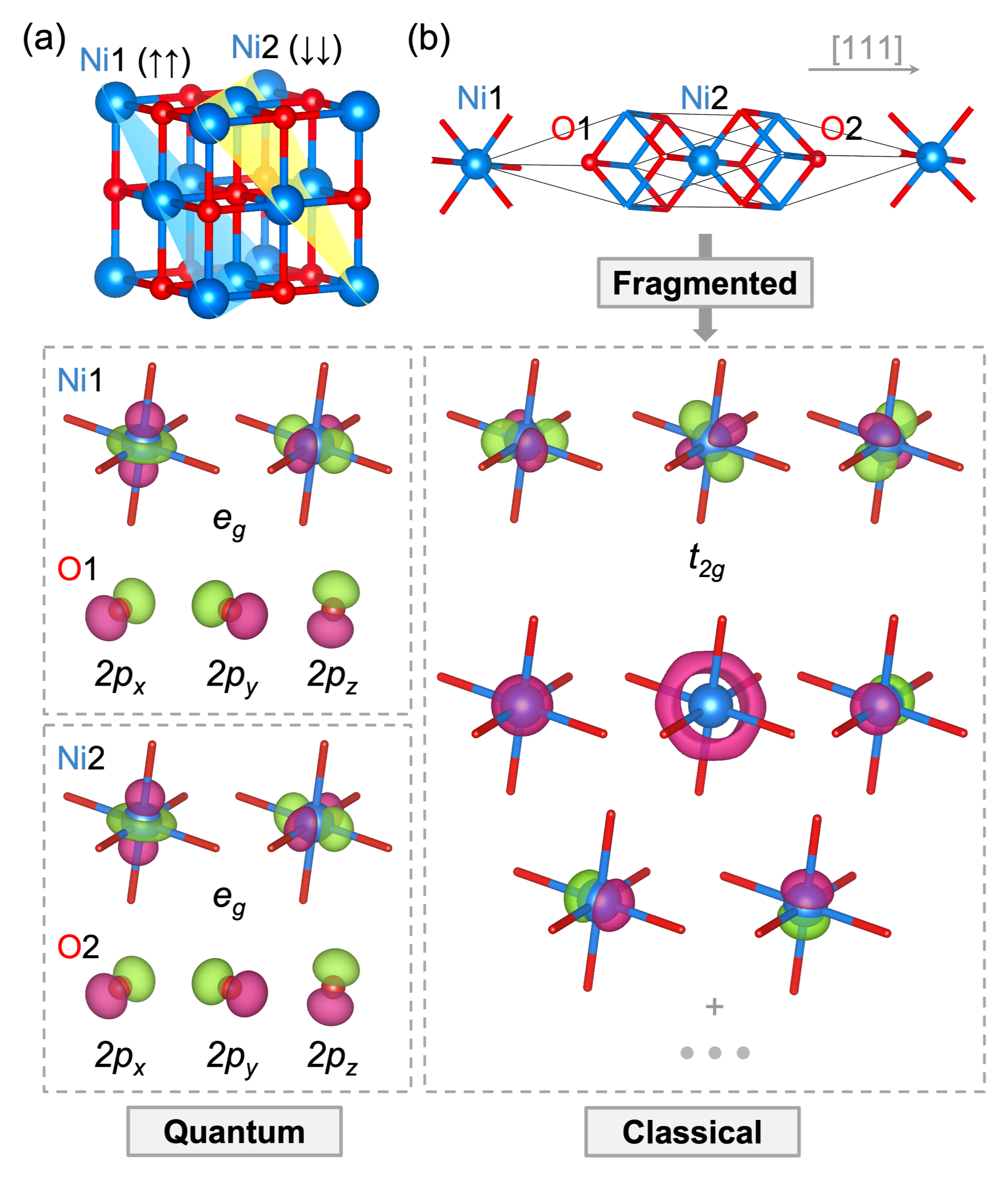}
  \caption{(a) Rock-salt structure of NiO with two ferromagnetic planes (blue and yellow) stacking antiferromagnetically. (b) The computational unit cell of NiO is fragmented into three fragments and their corresponding IAOs. The first and the second fragment are the $e_{g}$ and $2p$ orbitals from Ni$^{2+}$ and O$^{2-}$, respectively. The third fragment is the $t_{2g}$ of Ni$^{2+}$ and all other orbitals; only several of them are shown here.}
  \label{fig:NiO} 
\end{figure}

In this section, we further study the accuracy of our orbital-based partition approach using a high spin 3$d$ solid, NiO, a strongly correlated insulator whose electronic structure and magnetic properties have been extensively studied for decades~\cite{Graaf1997,Kodderitzsch2002,Hoffmann2020,Rak2020,Fischer2009,Chatterji2009,Balagurov2016,Keshavarz2018,Logemann2017,Booth2013}. It is well-documented that NiO crystallizes in a rock-salt structure with a type II antiferromagnetic ground state, \textit{i.e.}, an AFII state. First, all the magnetic moments localized on Ni$^{2+}$ in each (111) plane are ferromagnetically aligned. The ferromagnetic (111) planes are stacked antiferromagnetically along the [111] direction, resulting in the AFII state (see Fig.\ref{fig:NiO} (a)). If they are stacked ferromagnetically (all magnetic moments are aligned in the same direction), then we have an FM state of NiO. The energetic difference between these two states of NiO could be straightforwardly calculated using the nearest-neighbor and the next-nearest-neighbor exchange interaction, $J_1$ and $J_2$, respectively. The exchange interactions have been measured or calculated using various techniques, thereby providing reliable references to benchmark our calculations. We note that DMET and DMFT calculations using the unit cell(s) as the fragments have been reported on the AFII state of NiO in previous work~\cite{pDMET_Cui}.

In order to model the AFII and FM state of NiO, we use a rhombohedral structure with two formula units per unit cell. Within the octahedral crystal-field, the five-fold degenerate 3$d$ orbitals of Ni$^{2+}$ are split into $e_{g}$ and $t_{2g}$ orbitals. Inspired by the fact that the $t_{2g}$ orbitals are fully filled while the $e_{g}$ orbitals are half filled~\cite{Keshavarz2018} we partition the unit cell of NiO into three fragments as shown in Fig.~\ref{fig:NiO} (b). The first fragment contains the $e_{g}$ orbitals of the first Ni$^{2+}$ and the $2p$ orbitals of the first O$^{2-}$. We note that the inclusion of the $2p$ orbitals from O$^{2-}$ in this fragment is to account for the superexchange between Ni$^{2+}$ 3$d$ and O$^{2-}$ $2p$ orbitals~\cite{Graaf1997,Kodderitzsch2002}. The second fragment is composed of similar orbitals from the second Ni$^{2+}$ and O$^{2-}$. The third fragment is composed of all remaining orbitals which include the $t_{2g}$ orbitals as well as other orbitals with high angular momentum quantum numbers. Similar to h-BN, our partition scheme permits a mixed solver scheme where the QC solver is used for the two small fragments which presumably account for the static correlation while the classical CCSD solver is used for the large fragment to recover the remaining dynamic correlation. The resulting solver is denoted as QC/QC/CCSD. We also perform DMET calculations using the entire unit cell as the fragment to compare with our orbital-based partition scheme.

Table~\ref{table:NiO} presents our computed energy gaps between the FM and AFII states using DMET from two different partition schemes and solvers. First, we find that o-DMET tends to underestimate the gaps between FM and AFII and does not significantly differ from the FM-AFII gap computed by HF. At the 2$\times$2$\times$2 \textbf{k}-mesh, o-DMET using the unit cell as the fragment even incorrectly predicts FM to be the ground state of NiO. The sc-DMET predictions are in better agreement with the existing computational and experimental FM-AFII gaps from the literature. {Although, the importance of correlation potential fitting in DMET calculations for the electronic structure of magnetic NiO is highlighted by the improved results obtained with multi-fragment embedding, it's worth noting that the impact of self-consistency in DMET calculations can vary depending on the system and the flavor of DMET used. For instance, previous studies have reported minimal impact on single-embedding calculations of molecules\mbox{~\cite{CASDMET}} and Gamma-point solid-state systems\mbox{~\cite{mitra2021, DMET_CO}}, while improved results have been observed with multi-fragment embedding\mbox{~\cite{pDMET_Cui,pDMET_Hung,DMET_Knizia2}}}. Second, we find that the orbital-based partition scheme is overall in excellent agreement with the unit cell embedding despite employing much smaller fragments. At the 2$\times$2$\times$2 \textbf{k}-mesh, the energy gaps computed by the orbital-based sc-DMET are only around 20.0 meV (or 0.5 kcal/mol) different from that by \textbf{k}-CCSD, validating the accuracy of the orbital-based partition for calculating the FM-AFII gaps of NiO. Since there are precisely twenty spin orbitals within the embedding problem for each small fragment, we can employ VQE with twenty qubits as a DMET solver. The agreement of the FM-AFII gap calculated by our QC-DMET method with the classical results is within 40 meV, which falls within the chemical accuracy of 1 kcal/mol.

{Next, we show how the many-body screening effects between the fragments are recovered through the self-consistent optimization of the correlation potential by conducting a classical experiment in which we explicitly group the first two fragments together as one to explicitly account for the many-body screening effect between these two fragments. Specifically, we partition the unit cell into two fragments: (1) Ni$^{2+}$ $e_{g}$ orbitals and O$^{2-}$ $2p$  orbitals; (2) the remaining orbitals on Ni$^{2+}$ and O$^{2-}$. This experiment is denoted as CCSD/CCSD in Table~{\ref{table:NiO}}. Our results indicate that at the o-DMET level, the difference between the two methods of partitioning the unit cell (into two or three fragments) is insignificant at all \textbf{k}-meshes and at TDL. This suggests that the many-body screening effect between the two fragments is of minimal significance. Additionally, at the sc-DMET level, the energy gap computed by CCSD/CCSD (two-fragments) is significantly improved from that computed by CCSD/CCSD/CCSD (three-fragments). At $2^3$ \textbf{k}-mesh, the difference in the energy gap computed by CCSD/CCSD and \textbf{k}-CCSD is approximately 12.6 meV or 0.3 kcal/mol, which is smaller than the chemical accuracy. This suggests that the many-body screening effect between the fragments in our method is partially included and can be systematically improved by including more orbitals, as previously discussed. Thus, this experiment demonstrates that the many-body screening effect between fragments in our method can be partially included, and systematically improved by increasing the number of orbitals.}

We emphasize that we do not attempt to prove a quantum advantage within this approach, rather we demonstrate how the orbital-based partition for DMET could be utilized in conjunction with quantum solvers to study realistic and challenging solids. As far as we are concerned, this is one of the first examples, if not the first, of the \textit{ab initio} quantum computation on strongly correlated solids containing 3$d$ transition metal using a medium-sized basis set. Our proposed approach will benefit from both hardware and algorithmic advances from the quantum computing community.

\begin{table}[h]
\caption{The energy differences between the FM and AFII state ($\text{E}_\text{FM} - \text{E}_\text{AFII}$ in meV/formula unit) for NiO from different methodologies. Both one-shot (o-DMET) and self-consistent DMET (sc-DMET) are presented. For rows labeled with full cell CCSD, the whole unit cell is treated as the fragment while the others are calculated with multi-fragment partition. {The FM-AFII energy gaps were extrapolated to the TDL using $3^3$ and $4^3$ {\bf k}-point, which has been reported to give good agreement with experimental measurements, as demonstrated by McClain \textit{et al}.~\mbox{\cite{McClain2017}}}. The calculated and experimental values from the literature are also given.}
\footnotesize
\begin{tabular*}{1.0\linewidth}{p{1.45cm}cccccc}
\toprule
\multirow{2}*{Method} & \multirow{2}*{Solver} & \multicolumn{3}{c}{\textbf{k}-mesh} & {Extrap.} \\ 
 &  & $2^3$ & $3^3$ & $4^3$ & TDL \\ 
\midrule
HF & & 40.7	& 24.0 & 22.9 & 19.3  \\
o-DMET & Full cell CCSD & -23.4 &	49.6 & 46.1 & 35.5 \\
& CCSD/CCSD/CCSD & 95.8 & 33.6 & 30.3 & 20.4 \\
& FCI/FCI/CCSD & 97.5 & 34.6 & 31.4 & 22.0 \\
& QC/QC/CCSD & 87.9 & 33.4 & 31.2 & 24.3 \\
& QC/QC/HF & 54.5 & 33.3 & 31.3 & 25.3 \\
& CCSD/CCSD & 95.1 & 33.6 & 30.4 & 20.5 \\
\midrule
sc-DMET & Full cell CCSD & 95.4 & 51.9 & 57.4 & 73.8 \\
& CCSD/CCSD/CCSD & 109.0 & 43.0 & 43.3 & 44.4 \\
& FCI/FCI/CCSD & 111.7 & 45.6 & 52.8 & 74.2 \\
& QC/QC/CCSD & 94.3 & 25.4 & 26.7 & 30.5 \\
& QC/QC/HF & 57.7 & 34.8 & 36.9 & 43.0 \\
& CCSD/CCSD & 153.9 & 87.3 & 87.3 & 87.5 \\
\midrule
\textbf{k}-CCSD &  & 141.3 &   &   &  \\
\toprule
Method & \multicolumn{6}{c}{Literature}  \\
\midrule
HF  & \multicolumn{6}{c}{22.8\cite{Moreira2002}} \\
DFT & \multicolumn{6}{c}{89.9\cite{Keshavarz2018}, 100.8\cite{Moreira2002}, 103.7\cite{Rak2020},104.0\cite{Twagirayezu2019}, 111.1\cite{Keshavarz2018}, 145.8\cite{Moreira2002}} \\
SIC & \multicolumn{6}{c}{66.4\cite{Fischer2009},81.2\cite{Fischer2009}, 106.4\cite{Kodderitzsch2002}, 116.0\cite{Kodderitzsch2002}} \\
Experiment & \multicolumn{6}{c}{105.84\cite{Hutchings1972}, 112.2\cite{Shanker1973}} \\
\bottomrule
\label{table:NiO}
\end{tabular*}
\end{table}

\subsubsection{Quantum Resource Estimation.~~}
We highlight the computational efficiency of our framework by analyzing the required qubits for each system studied in this work (see Table \ref{table:qubits_survey}). One can easily see that the required number of qubits by a naive quantum algorithm would quickly explode as the dimension of the solids increases from 1D to 2D to 3D. For example, the largest calculation performed in this work on NiO with a 4 $\times$ 4 $\times$ 4 \textbf{k}-mesh would require 9984 qubits, which is far beyond the current status of any near-term quantum devices.

\begin{table}[h]
\small
  \caption{\ A survey on the number of required qubits for the largest simulation of each system studied in this work}
  \label{table:qubits_survey}
\footnotesize
  \begin{tabular*}{0.48\textwidth}{@{\extracolsep{\fill}}lccccc}
    \hline
    System  & Number of cGTOs & \textbf{k}-mesh &  \multicolumn{2}{c}{Number of Qubits} \\
&(per unit cell)&   & No embedding & This work \\
    \hline
    1D H    & 2  & $11^1 = 11$ & 44 & 8 \\
    2D h-BN & 26 & $7^2 = 49$  & 2548 & 12 \\
    3D NiO  & 78 & $4^3 = 64$  & 9984 &  20 \\
    \hline
  \end{tabular*}
\end{table}

\section{Conclusions}
In this work, we presented a flexible hybrid quantum-classical framework for quantum simulation of solid-state materials. With the multi-fragment partition of the unit cell at the orbital level, we show the possibility of \textit{ab initio} simulation of complex electronic structures on quantum computers. {Our approach is akin to the active space methods of quantum chemistry in that it involves dividing the unit cell into smaller fragments, similar to how an active space is partitioned into subspaces. However, there are several crucial distinctions in our method when compared to the traditional active space decomposition methods. Our proposed method represents a significant departure from the traditional active space decomposition methods such as Generalized Active Space Self-Consistent Field method~{\cite{GAS1,GAS2}}, Localized Active Space Self-Consistent Field method~{\cite{LAS1,LAS2}} or Active Space Decomposition~{\cite{ASD}}, which have been primarily formulated for molecular systems. In contrast, our method is formulated and implemented within a periodic DMET framework, enabling quantum computation on solid-state systems. Furthermore, our method diverges from previously studied multi-fragment DMET approaches, which have focused on weakly-bound fragments, by strategically partitioning fragments based on chemical intuition of orbital interactions. This results in a more flexible approach and the ability to control the size of the problem for near-term quantum devices.}

To investigate the accuracy of our method, we studied the antiferromagnetic electronic structure of the 1D hydrogen chain within periodic boundary conditions where the whole unit cell is chosen to be a correlating group of orbitals. This can be seen as a special case of our approach. The agreement between the EOS produced by our method and other classical approaches like \textbf{k}-CCSD or CCSD-DMET demonstrated the accuracy of the unrestricted UCCSD \textit{ansatz} used in DMET. We then calculated the EOS of h-BN and compare them against those calculated by \textbf{k}-CCSD. A good quantitative agreement was observed, indicating the accuracy of our method for this 2D material. The strongly correlated electronic structure of NiO was also studied. Not only do our calculations agree with the \textbf{k}-CCSD at the available \textbf{k}-mesh, but are also well consistent with previous studies from both experimental and theoretical literature.

The spin unrestricted UCCSD \textit{ansatz} was implemented and employed in the VQE procedures for 1D hydrogen chain and 3D NiO system, in particular, both occupations and orbital coefficients are not restricted to be the same for spin-up and spin-down electrons. Previous studies have focused extensively on the spin restricted wave function and the spin unrestricted \textit{ansatz} has been more or less overlooked. The spin unrestricted UCCSD solver could be useful for cases when it is more convenient to break the spin symmetry to converge to the ground state. We anticipate that the \textit{ansatz} improvement within this direction, for instance, an unrestricted $k$-UpCCGSD \textit{ansatz}~\cite{lee2018generalized} or an unrestricted adaptive variational algorithm~\cite{grimsley2019adaptive,huangEfficientQuantum2022} may further reduce the circuit depth with a satisfactory accuracy.

We anticipate several directions in which the recent technical advances that can be augmented with our framework to improve both accuracy and efficiency when dealing with large systems. One direction relates to the systematically improvable embedding scheme. In this scheme, the conventional bath orbitals in DMET can be systematically augmented with some extra orbitals computed from local second-order Møller‐Plesset perturbation~(MP2) calculations, resulting in a larger correlation space and higher accuracy~\cite{nusspickel2022systematic}. Another promising direction is to explore the use of quantum-classical hybrid quantum Monte Carlo~(QMC) solvers, for examples, a quantum-classical hybrid full configuration interaction QMC~\cite{zhang2022quantum} or auxiliary field QMC~\cite{huggins2022unbiasing}. These QMC enhanced methods are based on the fact that quantum computers could provide a better initial trial state with good overlaps with the ground state of the system. {Furthermore, incorporating higher level excitations in UCC \textit{ansatz} may lead to more accurate results, but currently requires a large number of parameters and deep quantum circuits. With advancements in parameter reduction and circuit compilation techniques, it may become practical\mbox{\cite{haidarExtensionTrotterized2022}}. }

Finally, we propose two applications that our method could potentially apply to. First, the correlating orbitals embedding studied herein could be used to study high-temperature superconducting cuprate on a NISQ device owing to the fact that Cu$^{2+}$ 3$d$ and O$^{2-}$ $2p$ orbitals govern the strong correlation physics in cuprate similar to Ni$^{2+}$ 3$d$ and O$^{2-}$ $2p$ orbitals in the case of NiO. We note that a comprehensive DMET study on the parent state of cuprate has been recently reported by Cui \textit{et al.}~\cite{cui2021systematic}. Thus, it would be an interesting complementary study on these materials using the orbital-based multi-fragment embedding proposed in this work. Solid-state heterogeneous catalysis like single atom/cluster catalyst~\cite{Qiao2011,maSurfaceSingleCluster2018} is another important chemistry application where the 3$d$ orbitals of transition metals play a significant role in determining the reaction barriers, thereby requiring an accurate treatment for electron correlations. In both cases, the application of our methodology requires a priori partition of different groups of active orbitals using chemical intuition. Our recommended practice is to systematically expand each correlating fragment by adding orbitals from the weakly correlated fragment until the property of interest, \textit{e.g.} relative energy, is converged.

\section*{Data availability}
{Data and template input files for this work can be obtained on Zenodo at https://doi.org/10.5281/zenodo.7514998. The source code is available upon reasonable request.}

\section*{Author contributions}
D. Lv designed the project. 
C. Cao, H. Q. Pham and D. Lv implemented the algorithm and collected the data. 
All authors prepared the manuscript.

\section*{Conflicts of interest}
The authors declare that there are no competing interests.

\section*{Acknowledgements}
The authors gratefully thank Zhihao Cui, Xiaojie Wu, Yi Fan, Xiang Li, Yifei Huang, Huanhuan Ma, Wei Hu, Qiming Sun, Xuelan Wen, Zhendong Li, Nan Sheng and Ji Chen for helpful discussions and Hang Li for support and guidance.


\bibliography{main.bib} 

\end{document}


\title{Supporting Information for ``\textit{Ab initio} Quantum Simulation of Strongly Correlated Materials with Quantum Embedding''}

\newcommand{\bytedance}{ByteDance Inc, Zhonghang Plaza, No. 43, North 3rd Ring West Road, 100098 Beijing, China.}
\newcommand{\bytedanceus}{ByteDance Inc, Mountain View, CA 94041, US.}
\newcommand{\tsinghua}{Department of Chemistry, Tsinghua University, Beijing 100084, China}
\newcommand{\cas}{Institute of Computing Technology, Chinese Academy of Sciences}
\newcommand{\ucas}{University of Chinese Academy of Sciences}
\newcommand{\peking}{Center on Frontiers of Computing Studies, Peking University, Beijing 100871, China}

\newcommand{\oxford}{Clarendon Laboratory, University of Oxford, Oxford OX1 3PU, United Kingdom}

\author{Changsu Cao}
\affiliation{\bytedance}
\affiliation{\tsinghua}

\author{Jinzhao Sun}
\affiliation{\oxford}

\author{Xiao Yuan}
\affiliation{\peking}
\affiliation{School of Computer Science, Peking University, Beijing 100871, China}

\author{Han-Shi Hu}
\affiliation{\tsinghua}

\author{Hung Q. Pham}
\email{hung.pham@bytedance.com}
\affiliation{\bytedanceus}

\author{Dingshun Lv}
\email{lvdingshun@bytedance.com}
\affiliation{\bytedance}

\maketitle

\newpage
\section{Computational Details}
The DMET workflow was implemented based on the package libDMET~\cite{pDMET_Cui,zhuEfficientFormulation2020}.
The classical calculations of FCI, CCSD and HF were performed by using PySCF in this work~\cite{PySCF_1,PySCF_2}. GTH pseudopotentials were employed to replace core electrons~\cite{GTHpotential1,GTHpotential2}. The GTH-SZV basis set was used to represent the valence electron of H~(1s) while GTH-DZVP-MOLOPT-SR basis sets were used for B, N, O, and Ni~\cite{GTHbasis}. These correspond to 2s2p3s3p3d AOs for B, N, O and 3s3p3d4s4p4d4f5s AOs for Ni. Gaussian density fitting~(GDF) was applied to evaluate the two-electron integrals~\cite{GDF}. An even-tempered Gaussian basis set was used as the auxiliary basis set with the exponential factor $\beta=2.0$ for 1D-H and $\beta=2.3$ for 2D-hBN and 3D-NiO~\cite{GDFAuxBasis}. 
{The quantum simulations of embedding were performed by virtual quantum simulators on classic devices. The detailed implementation can be found according to Data Availability section.} Yao~\cite{Yao} was used as the backend of QC solver when simulating 1D-H and 2D-hBN. For the larger simulations of NiO, Fermionic Quantum Emulator~(FQE)~\cite{FQE} was chosen as the backend of QC solver to speed up the numerical calculations. {The noisy simulations were based on the QASM simulator from Qiskit toolkit{~\cite{Qiskit}}. To mitigate quantum noise, the quantum observable is measured at noise-scaled quantum circuits by unitary folding and extrapolated to the zero-noise limit by linear fit supported by Mitiq package{~\cite{Mitiq}}. The scaling factors are 1.00, 1.25 and 1.50 in this work.
}

The momentum space was sampled using a $\Gamma$-centered Monkhorst-Pack grid.
For 1D-H, 10 \rAA$ $ vacuum region was set to avoid interactions between the neighboring layers due to periodic boundary conditions. A 1$\times$1$\times$N$_k$ \textbf{k}-mesh with N$_k$ = 3, 5, 7, 9, 11 was studied. Similarly for h-BN, a 20.0 \rAA$ $ vacuum region was used to model the single layer structure. We employed three expanding \textbf{k}-meshes, \textit{i.e.}, 5$\times$5$\times$1, 6$\times$6$\times$1, and 7$\times$7$\times$1. For NiO, both FM and AFII states are calculated using the lattice parameter $a=4.17 $ \rAA$ $ from the experiment~\cite{lattice_NiO_1983}. The \textbf{k}-point of 2$\times$2$\times$2, 3$\times$3$\times$3 and 4$\times$4$\times$4 were employed, however, only the latter two were used to extrapolate the energy gaps to the thermodynamic limit~(TDL). This two-point TDL extrapolation was reported to give good agreements with experimental measurements on silicon and carbon crystal~\cite{McClain2017}. For a non-embedding reference, we were only able to perform the \textbf{k}-sampled periodic CCSD (\textbf{k}-CCSD) calculation at the 2$\times$2$\times$2 \textbf{k}-point due to the current limitation of the computational resources.

The spin density in Fig.~2 is populated in Mulliken way~\cite{mullikenElectronicPopulation1955} that the off-diagonal terms in the one-particle reduced density matrix is equally attributed to the corresponding orbitals. The spin density of atom $x$ is defined as difference of spin-up and spin-down electron density of atom $x$, which is expressed as

\begin{equation}
\rho_{x} = \sum_{i \subset x}{(D_{ii,\alpha} - D_{ii,\beta})} + \frac{1}{2}\sum_{i \subset x,j \not\subset x }{(D_{ij,\alpha} - D_{ij,\beta})},
\label{SpinDensity}
\end{equation}

where $D$ is the one-particle reduced density matrix. 

\section{Extrapolation to Thermodynamic Limit of 1D-H Chain }
{We performed DMET calculations of 1D hydrogen chain with different k-meshes, ranging from 1x1x5 to 1x1x45. Both the antiferromagnetic state~(AFM) and the non-spin polarized state~(NSP) have been calculated. The bond distance that the ground state transferring from NSP to AFM is shown in Fig.~{\ref{fig:ToTDL_H2}}.

\begin{figure}[htp!]
    \centering
        \includegraphics[width=1.0\linewidth]{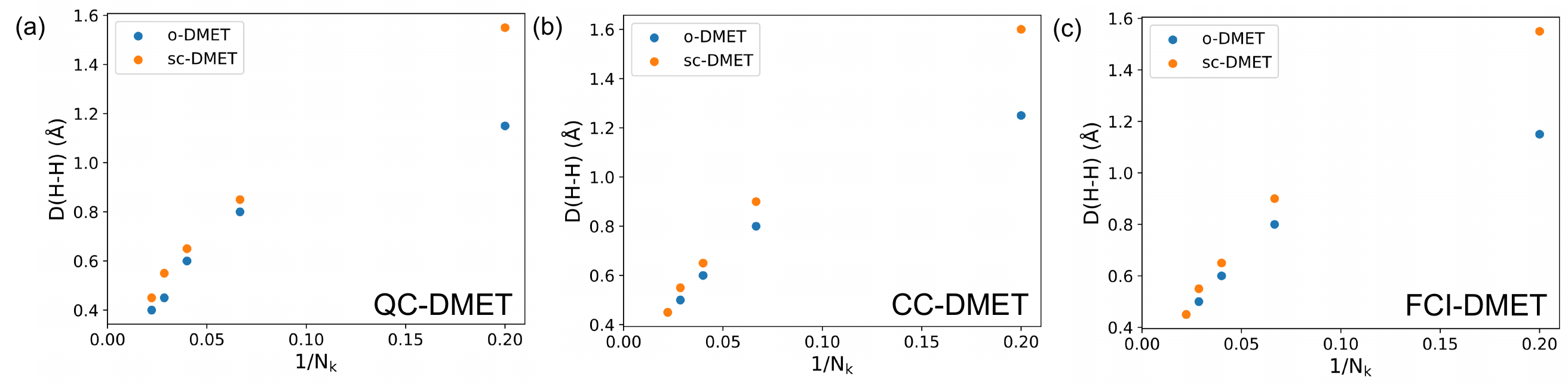}
        \caption{The reciprocal k-point number vs. bond distance at which the ground state changes, as calculated by three different methods: (a) QC-DMET, (b) CC-DMET, and (c) FCI-DMET.}
        \label{fig:ToTDL_H2}
\end{figure}
According to Fig.~{\ref{fig:ToTDL_H2}}, the DMET calculation with solvers predict similar bond distances of phase transition. We find that the predicted bond distance of sc-DMET is slightly larger than that of o-DMET. Our results are in agreement with previous studies, such as Liu et al.~{\cite{JCTC_HChain}} and Motta et al.~{\cite{HChain_PRX}}. As a detailed example, we focus on the result of QC-DMET. Liu et al.~{\cite{JCTC_HChain}} predicted the AFM state transfers at the range of 0.5-0.75 $\AA$ via AFQMC on hydrogen chain with the total number of atom N=50, which is consistent with the sc-DMET results~(0.65 $\AA$, $N_k=1/25$). Similarly, Motta et al.~{\cite{HChain_PRX}} predicted the phase transition at 0.85 $\AA$ using AFQMC, VMC and DMC with number of hydrogen atoms N=40, which is, however, slightly larger than the value, $ca.$ 0.69 $\AA$ ($N_k=1/20$), from the linear extrapolation of sc-DMET in Fig.~{\ref{fig:ToTDL_H2}}(a). Overall, our results demonstrate consistency with previous studies.}

\section{The energy differences of 1D-H}

\begin{figure}[htp!]
        \centering
          \includegraphics[width=1.0\linewidth]{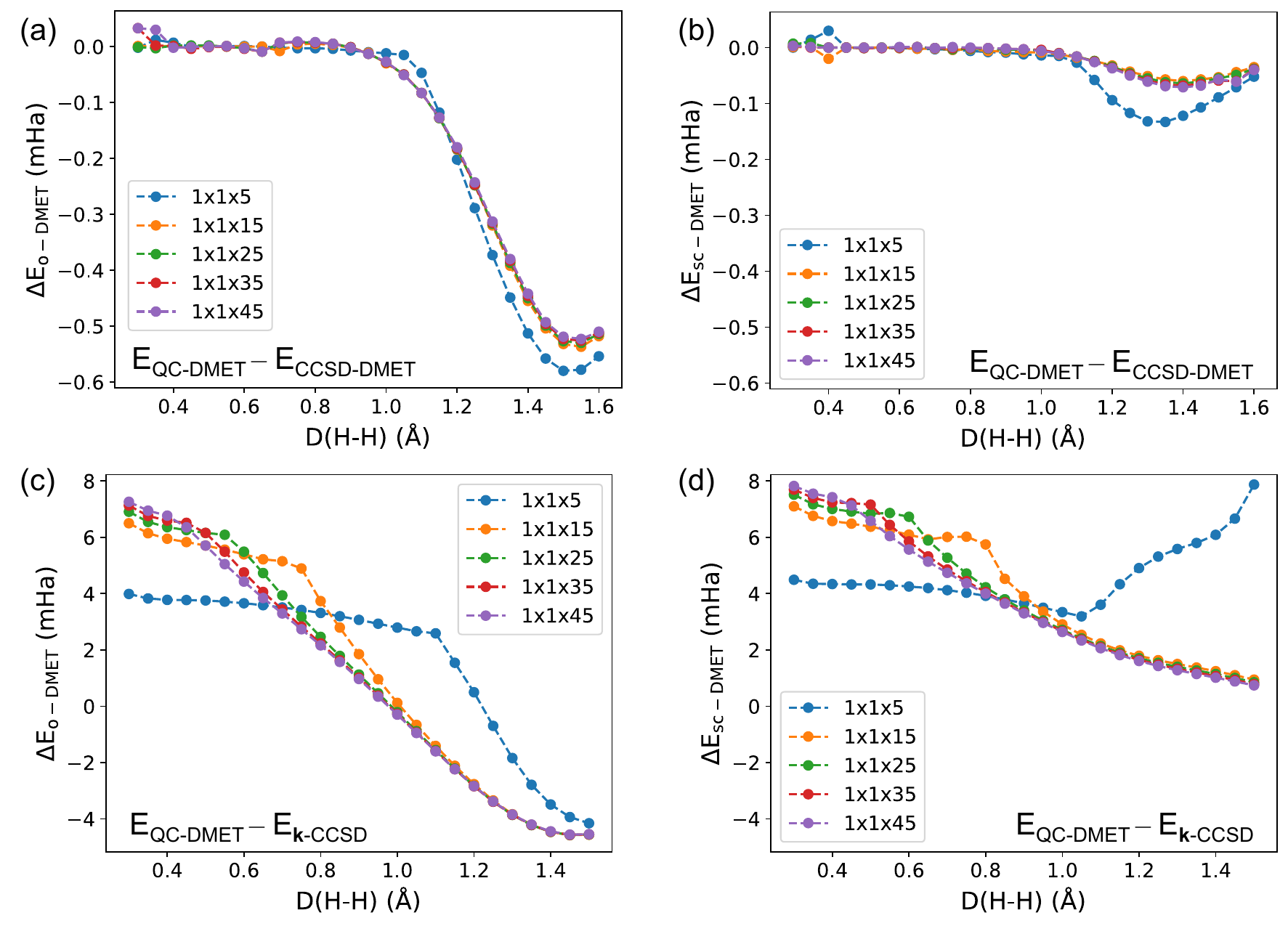}
          \caption{
          Energy differences of (a) o-DMET (QC solver vs CCSD solver), (b) sc-DMET (QC solver vs CCSD solver), (c) o-DMET (QC solver) vs k-CCSD (no embedding), and (d) sc-DMET (QC solver) vs k-CCSD (no embedding) calculations, plotted against H-H bond distance. Ground state of each dot in figure refers to result of QC-DMET as different methods predict different phase transition bond distance. The same scale of vertical axis is used for the sub-figures in the same row.
          }
          \label{fig:Ediff_H2_PES}
\end{figure}

{To determine the accuracy and reliability of our QC solver, we compare its energy differences with other results in Fig.~{\ref{fig:Ediff_H2_PES}}. We find that the energy difference between our QC solver and CCSD solver is less than 1 mHa for all bond distances studied. For both o-DMET and sc-DMET, the energy difference increases as the bond dissociates. The energy difference for sc-DMET is particularly small in the dissociation zone. The agreement between our QC solver and CCSD solver improves as the k-mesh becomes denser. When comparing with $\bf{k}$-CCSD results without embedding (shown in Fig.~{\ref{fig:Ediff_H2_PES}}), we find that the deviation caused by embedding is larger than the energy difference among different solvers used in the DMET calculation. For o-DMET with our QC solver, the energy is several milli-Hartrees higher than the results of $\bf{k}$-CCSD in the compressed zone and lower in the more stretched zones. After incorporating correlation potential fitting (sc-DMET), we observe that the energy difference in the compressed region remains small, but is significantly reduced by the self-consistent fitting of the correlation potential, particularly around the stretched region where D(H-H) $\geq$ 1.0 \AA.}

\section{The energy differences between the FM and AFII state for NiO} 

The exchange coupling constants $J_1$ (nearest-neighbor) and $J_2$ (next-nearest-neighbor) for NiO are defined as~\cite{Kodderitzsch2002}

\begin{equation}
J_1 = \frac{1}{16}(\text{E}_{\text{AFI}} - \text{E}_{\text{FM}}),~~
J_2 = \frac{1}{48}(4\text{E}_{\text{AFII}} - 3\text{E}_{\text{AFI}} - \text{E}_{\text{FM}})
\label{J2}
\end{equation}

where $\text{E}_{\text{AFI}}$, $\text{E}_{\text{AFII}}$, and $\text{E}_{\text{FM}}$ are the energy of the AFI, AFII, and FM state, respectively. Thus, the energy differences between the FM and AFII state can be calculated as

\begin{equation}
\text{E}_\text{FM} - \text{E}_\text{AFII} = -12 \times (J_1 + J_2)
\label{FM_AFII}
\end{equation}

\begin{table}[h]
\begin{center}
\caption{The energy differences between the FM and AFII state ($\rm{E_{FM}} - \rm{E_{AFII}}$ in meV/formula unit) for NiO from the literature.}
\begin{tabular}{ccccc}
\toprule
Method & Ref. & $J_1$ & $J_2$ & $\text{E}_\text{FM} - \text{E}_\text{AFII}$\\
\midrule
\hline
Exp. & \cite{Shanker1973} & -0.69 & -8.66 & 112.2 \\
Exp. & \cite{Hutchings1972} & 0.69 & -9.51 & 105.8 \\
HF & \cite{Moreira2002} & 0.4 & -2.3 & 22.8 \\
LSDA+U & \cite{Keshavarz2018} & -0.03 & -7.46 & 89.9 \\
Fock35 & \cite{Moreira2002} & 0.95 & -9.35 & 100.8 \\
GGA+U & \cite{Twagirayezu2019} & 0.87 & -9.54 & 104.0 \\
LDA+U & \cite{Keshavarz2018} & 0.01 & -9.27 & 111.1 \\
PBE & \cite{Rak2020}	 & 0.83 & -9.47 & 103.7 \\
B3LYP & \cite{Moreira2002} & 1.2  & -13.35  & 145.8 \\
SIC-LMTO & \cite{Kodderitzsch2002} & 0.9  & -5.5  & 55.2 \\
LSIC-LSDA & \cite{Fischer2009} & 0.15 & -6.92  & 81.2 \\
LSIC-LSDA & \cite{Fischer2009} & 1.42 & -6.95 & 66.4 \\
SIC-LSDA & \cite{Kodderitzsch2002}  &  &  & 116.0 \\
SIC-LSDA (ES) & \cite{Kodderitzsch2002}  &  &  & 106.4 \\
\hline
\bottomrule
\label{nio_gaps}
\end{tabular}
\end{center}
\end{table}

\section{Quantum Noisy Simulations}
{The noisy quantum simulations are performed by introducing the effect of measurement shots and depolarizing noise to evaluate the performance of this method in more realistic conditions.}

{The noisy simulations are performed on closed-shell 1D hydrogen chain with GTH-SZV basis set. The distance between adjacent hydrogen atom is 1.5 \AA. We choose a minimum fragment which contains one hydrogen atom due to the expensive computational cost when including measurements. The corresponding embedding thus contains 4 spin-orbitals and has the size of 4 qubits. Only chemical potential fitting is performed here.}
\begin{figure}[htp!]
\centering
\includegraphics[width=1.0\linewidth]{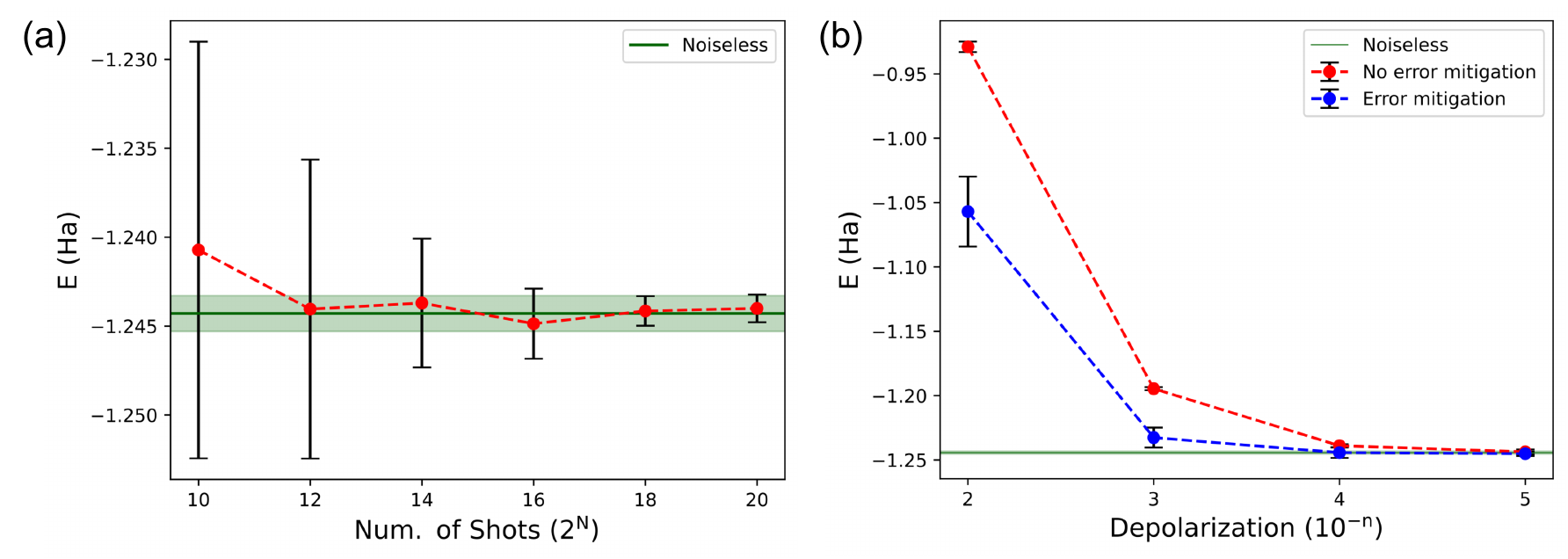}
\caption{(a) DMET energies vs. number of shots without extra gate noise. (b) DMET energies vs. depolarization probability with $2^{18}$ shots. The green line represents the noiseless reference calculated by numerical ideal solvers. The zone where the energy difference from the noiseless value is within 1.0 mHa has been marked by light green. Numerical experiments are repeated 10 times for each dot to get the mean and standard deviation.}
\label{fig:Noisy_Results}
\end{figure}

{We first consider only the fluctuation caused by finite shots without any gate noise involved. Here one shot represents a single execution of the quantum algorithm. Simulated results with the number of shots ranging from $2^{10}$ to $2^{20}$ are depicted in Fig.{~\ref{fig:Noisy_Results}}(a). The precision of the DMET energy improves and converges well as the increase of number of shots. The standard deviation becomes smaller than 1.0 mHa when the number of shots reaches $2^{18}$. Since no extra gate noise is introduced, the mean value of DMET energies shows high accuracy and the difference from the ideal reference value is smaller than 1.0 mHa even with several thousands shots ($\sim 2^{12}$).}

{In the following, the depolarization noise is introduced in the DMET simulation. The depolarization noise is one of the most commonly used noise model in quantum simulations. By applying the depolarization noise model, there may be a bit-flip~(Pauli $\bf{X}$ error) or a phase-flip~(Pauli $\bf{Z}$ error) or both~(Pauli $\bf{Y}$ error) to the input state based on the depolarizing probability $p$, when executing an operation in the quantum circuit~{\cite{babarDualityQuantum2019,terhalQuantumError2015}}.}
{As shown in Fig.~{\ref{fig:Noisy_Results}}(b), the depolarizing noise leads to less accurate DMET energies while the standard deviation is just slightly affected by it. By applying the zero-noise extrapolation of linear fit, the more accurate results can be obtained at the expense of larger standard deviation. This worse precision after error mitigation is attributed to the deeper quantum circuit generated in the zero-noise extrapolation process.}

{In summary, the measurement shot mainly affects the precision while the depolarizing noise affects more the accuracy of the DMET results. The zero-noise extrapolation based on linear fit, which is one of the simplest error mitigation strategy, is used here to mitigate the quantum error. It improves the accuracy of the DMET energy while sacrificing the precision to some degree. It is a similar optimization procedure of chemical potential fitting, but more complex. We expect a good convergence achievable provided more robust error mitigation strategies and advanced sampling methods. Some recent studies show that the self-consistent quantum-classical hybrid algorithms remain well converged in the presence of noise~{\cite{liuBootstrapEmbedding2023,tillyReducedDensity2021}}. Quantum error mitigation strategies such as zero-noise extrapolation~{\cite{temmeErrorMitigation2017,liEfficientVariational2017}}, stochastic methods~{\cite{wallmanNoiseTailoring2016,sunMitigatingRealistic2021}} and subspace expansion~{\cite{bonet-monroigLowcostError2018}.} Besides that, reducing the depth of quantum circuit based on molecular symmetry~{\cite{C2H4_28qubits}} or in an adaptive manner~{\cite{grimsley2019adaptive,fanCircuitDepthReduction2021}} can help to sample the observable with noise presented. Grouping commute Pauli terms can also reduce the cost of measurement for easier measurement~(for example tensor product basis~{\cite{bravyiTaperingQubits2017}}.}

\bibliography{si.bib}